\documentclass[a4paper,12pt,english]{article}
\usepackage[english]{babel}
\usepackage{amsmath}
\usepackage{amsfonts}
\usepackage{amssymb}
\usepackage{geometry}
\usepackage{authblk}
\usepackage{enumitem}

\setlist{leftmargin=1.5cm}

\newcommand{\ie}{{\it i.e.,} }

\makeatletter
\def\@maketitle{%
\newpage
\null
\begin{flushright}
FIAN/TD/17-15
\end{flushright}
\vskip 2em%
\begin{center}%
\let \footnote \thanks
{\LARGE \@title \par}%
\vskip 1.5em%
{\large

\lineskip .5em%
\begin{tabular}[t]{c}%
\@author
\end{tabular}\par}
\vskip 1em%
{\large \@date}%
\end{center}
\par
\vskip 1.5em
}
\makeatother

\begin{document}

\thispagestyle{empty}
\title{\textbf{Gauge Non-Invariant Higher-Spin Currents in $AdS_4$}}
\author{\textbf{P.A. Smirnov and M.A. Vasiliev}}
\affil{\small \emph{I.E.Tamm Department of Theoretical Physics, Lebedev Physical Institute of RAS,}\\
\emph{Leninsky prospect 53, 119991, Moscow, Russia}}
\date{}
\maketitle
\thispagestyle{empty}
\begin{abstract}
Conserved currents of any spin $t>0$ built from bosonic symmetric
massless gauge fields of arbitrary integer spins $s_1 +s_2 >t$ in $AdS_4$
are found. Analogously to the case of $4d$ Minkowski space,
currents considered in this paper are not
gauge invariant but generate gauge invariant
conserved charges.
\end{abstract}

\newpage
\setcounter{page}{1}
\section{Introduction}
Gauge-invariant conserved currents are well known and
were deeply studied in the literature
\cite{c2,c3,c4,c5,c6,Gelfond:2006be,c8,Kaparulin:2011aa,Gelfond:2014pja}.
In the general case, a conserved current carries a set of
three spins $(t, s_1, s_2)$, where $t$ is a spin of the
current itself, and $s_1$ and $s_2$ are spins of the fields it is
constructed from. For example, the so-called gravitational
stress pseudo-tensor \cite{LL} ($s=t=2$ conserved current)
is not gauge invariant.
The same fact is shown in
\cite{c6} for the $t=2$ current
built from massless fields of spins $s>2$.
The spin-zero field has no gauge symmetry; thus, the currents with $s < 1$
are gauge invariant, while the spin-one
current built from two massless spin-one fields is not.

The aim of this paper is to extend the Minkowski-space results of \cite{Smirnov:2013},
 presenting the full list of gauge non-invariant currents
with integer spins in $AdS_4$ such that $t< s_1 +s_2$. Being
gauge non-invariant, these currents give rise to the
gauge-invariant conserved charges. Gauge non-invariant currents
 will be derived from the variation
of the cubic action of \cite{Fradkin:1986qy, Fradkin:1987ks}, which is
gauge invariant in the lowest order.

\subsection{Conventions}
In this paper, we consider $AdS_4$ space-time.
Greek indices $\mu, \nu, \rho, \lambda, \sigma$
are the base and range from 0--3. Other Greek indices are spinorial and take values of one and two.
The latter are raised and lowered by the \emph{sp}(2)
antisymmetric forms:
$\varepsilon_{\alpha\beta}, \varepsilon^{\alpha\beta},
\varepsilon_{\dot\alpha\dot\beta}, \varepsilon^{\dot\alpha\dot\beta}$
\begin{gather}
\varepsilon^{\alpha \beta}\varepsilon_{\alpha \gamma} = \delta^\beta_\gamma, \quad
\varepsilon^{\dot\alpha \dot\beta}\varepsilon_{\dot\alpha \dot\gamma} = \delta^{\dot\beta}_{\dot\gamma}, \\
A_\alpha = A^{\beta}\varepsilon_{\beta\alpha}, \quad
 A^\alpha = A_{\beta}\varepsilon^{\alpha\beta}, \quad
A_{\dot\alpha} = A^{\dot\beta}\varepsilon_{\dot\beta\dot\alpha}, \quad
A^{\dot\alpha} = A_{\dot\beta}\varepsilon^{\dot\alpha\dot\beta}.
\end{gather}

 Complex conjugation $\bar A$ relates dotted and undotted spinors.
Brackets ($[...]$) $\{...\}$ imply complete (anti)symmetrization, \ie
\begin{equation}
A_{[\alpha}B_{\beta]}=\frac{1}{2}(A_\alpha B_\beta-A_\beta B_\alpha)\,,\qquad
A_{\{\alpha}B_{\beta\}}=\frac{1}{2}(A_\alpha B_\beta+A_\beta B_\alpha).
\end{equation}
$A^{\alpha(m)}$ denotes a totally symmetric
multispinor $A^{\{\alpha_1 \ldots \alpha_m\}}$.

The wedge symbol $\wedge$ is implicit.

\section{Fields, Equations, Actions}

In the four-dimensional case considered in this paper,
it is convenient to use the frame-like formalism in
two-component spinor notation.
In these terms, a bosonic spin-$s$ Fronsdal field
 \cite{Fronsdal: 1978} is represented by multispinor one-forms
 \cite{Vasiliev:1987fortschr}:

\begin{equation*}
s\geq1:\quad\varphi_{\mu_{1}...\mu_{s}}\rightarrow\{
\omega^{\alpha(m)}{}_{,}{}^{\dot\beta(n)}\mid m+n = 2(s-1)\}\,,
\quad \omega^{\alpha(m)}{}_{,}{}^{\dot\beta(n)}=dx^\mu \omega_{\mu}{}^{\alpha(m)}{}_{,}{}^{\dot\beta(n)},
\end{equation*}
which are symmetric in all dotted and all undotted spinor indices
and obey the reality condition \cite{Vasiliev:1987fortschr}:
\begin{equation}
\omega_{\alpha(m),\dot\beta(n)}^\dag = \omega_{\beta(n),\dot\alpha(m)} .
\end{equation}

The frame-like field is a particular connection at $n=m=s-1$ ($s$ is integer):
\begin{equation}
h_{\mu}{}^{\alpha(s-1)}{}_{,}{}^{\dot\beta(s-1)}dx^{\mu}
:= \omega_{\mu}{}^{\alpha(s-1)}{}_{,}{}^{\dot\beta(s-1)}dx^{\mu}\,.
\end{equation}
By imposing appropriate constraints, the
connections $\omega^{\alpha(m),\dot\beta(n)}$ can be
expressed via $t = \frac{1}{2} |m-n|$ derivatives of the frame-like field \cite{Vasiliev:1987fortschr}.

Background gravity is described by the vierbein
one-form $\tilde h^{\alpha}{}_{,}{}^{\dot\beta}$ and
one-form connections $\tilde\omega^{\dot\alpha\dot\beta}, \ \tilde\omega^{\alpha\beta}$.
Lorentz covariant derivative $\tilde D$ acts as usual:
\begin{equation}
\tilde D A^{\alpha(m)}{}_{,}{}^{\dot\beta(n)}=dA^{\alpha(m)}{}_{,}{}^{\dot\beta(n)}+
m \tilde \omega^{\alpha}{}_{\gamma}A^{\alpha(m-1)\gamma}{}_{,}{}^{\dot\beta(n)}
+n \tilde \omega^{\dot\beta}{}_{\dot\delta}A^{\alpha(m)}{}_{,}{}^{\dot\beta(n-1)\dot\delta}
\end{equation}
for any multispinor $A^{\alpha(m),\dot\beta(n)}$.
The torsion and curvature two-forms are:
\begin{gather}
\tilde R^{\alpha}{}_{,}{}^{\dot\beta}
=d{\tilde h}^{\alpha}{}_{,}{}^{\dot\beta}+
{\tilde\omega}^{\alpha}{}_{\gamma}{\tilde h}^{\gamma}{}_{,}{}^{\dot\beta}+
{\tilde\omega}^{\dot\beta}{}_{\dot\delta}{\tilde h}^{\alpha}{}_{,}{}^{\dot\delta}\,,\\
\tilde R^{\alpha\alpha}=d{\tilde\omega}^{\alpha\alpha}+
{\tilde\omega}^{\alpha}{}_{\gamma}{\tilde\omega}^{\alpha\gamma}+
\lambda^2 {\tilde h}^{\alpha}{}_{,\dot\delta}
\tilde h^{\alpha}_{}{,}^{\dot\delta}\,,\\
\tilde R^{\dot\beta\dot\beta}=
d{\tilde\omega}^{\dot\beta\dot\beta}+
{\tilde\omega}^{\dot\beta}{}_{\dot\gamma}{\tilde\omega}^{\dot\beta\dot\gamma}+
\lambda^2 {\tilde h}_{\gamma,}{}^{\dot\beta}
{\tilde h}^{\gamma}_{}{,}^{\dot\beta},
\end{gather}
where the parameter $\lambda$ is proportional to the inverse $AdS$ radius
$\lambda \sim r^{-1}$.
$AdS_4$ space is described by the vierbein and connections obeying the equations:
\begin{equation}
\label{min}
\tilde R^{\alpha}{}_{,}{}^{\dot\beta}=0,\qquad
\tilde R^{\alpha\alpha}=0, \qquad
\tilde R^{\dot\beta\dot\beta}=0.
\end{equation}

Linearized higher-spin (HS) curvatures are:
\begin{align}\label{LinHS}
&R_1{}^{\alpha(m)}{}_{,}{}^{\dot\beta(n)}=
\tilde D\omega^{\alpha(m)}{}_{,}{}^{\dot\beta(n)}+ n(\theta(m-n) +
\lambda^2\theta(n-m-2) )
\tilde h_{\gamma,}{}^{\dot\beta}\omega^{\gamma\alpha(m)}{}_{,}{}^{\dot\beta(n-1)}+\nonumber\\
&\qquad\qquad\qquad\qquad+m(\theta(n-m) + \lambda^2 \theta(m-n-2))
\tilde h^{\alpha}{}_{,\dot\delta}\omega^{\alpha(m-1)}{}_{,}{}^{\dot\beta(n)\dot\delta},
\end{align}
where $\theta(x)$ is the step-function:
\begin{gather}
\theta(x)=
\begin{cases}
1&\text{at $x \geq 0$;}\\
0&\text{at $x < 0$.}
\end{cases}
\end{gather}
Curvatures (\ref{LinHS}) obey the Bianchi identities \cite{Vasiliev:1987fortschr}:
\begin{multline}\label{Bianchi}
\tilde D R_1{}^{\alpha(m)}{}_{,}{}^{\dot\beta(n)} =
-\lambda^{(|m-n|/2) + 1}
(m\lambda^{-|m-n-2|/2}\tilde h^\alpha{}_{,\dot\delta}R_1{}^{\alpha(m-1)}{}_{,}{}^{\dot\beta(n)\dot\delta}+\\
+ n\lambda^{-|m-n+2|/2}\tilde h_{\gamma,}{}^{\dot\beta}R_1{}^{\alpha(m)\gamma}{}_{,}{}^{\dot\beta(n-1)}).
\end{multline}

It is convenient to introduce two-forms $H_{\alpha \beta}$ and $\bar H_{\dot\alpha\dot\beta}$:
\begin{gather}
\label{hhH}
\tilde h_{\alpha,}{}_{\dot\beta}\tilde h_{\gamma,}{}_{\dot\delta}=
\frac{1}{2}\epsilon_{\alpha \gamma} \bar H_{\dot\beta\dot\delta}+
\frac{1}{2}\epsilon_{\dot\beta\dot\delta} H_{\alpha \gamma}\,,\\
\label{H}
H_{\alpha \beta}:=\tilde h_{\alpha, \dot\gamma}
\tilde h_{\beta,}{}^{ \dot\gamma}\,,
\quad
\bar H_{\dot\alpha\dot\beta}:=\tilde h_{\gamma, \dot\alpha}
\tilde h^\gamma{}_{,\dot\beta}.
\end{gather}

Free field equations for massless fields of spins $s\geq 2$
in Minkowski space
can be written in the form \cite{Vasiliev:1987fortschr}:
\begin{align}
\label{EM}
&R_1{}^{\alpha(m)}{}_{,}{}^{\dot\beta(n)} = 0 \qquad \text{for} \qquad n>0, \ m>0, \ n+m = 2(s-1); \\
\label{EMCundot}
&R_1{}^{\alpha(m)} = C^{\alpha(m)\gamma\delta} \
H_{\gamma \delta}
\qquad \text{for} \qquad m=2(s-1);\\
\label{EMCdot}
&R_1{}^{\dot\beta(n)} =
 \bar C^{\dot\beta(n)\dot\gamma\dot\delta} \ \bar H_{\dot\gamma\dot\delta}
\qquad \text{for} \qquad n=2(s-1).
\end{align}
Equations (\ref{EM})--(\ref{EMCdot}) are equivalent to the equations of motion, which
follow from the Fronsdal action \cite{Fronsdal: 1978}
supplemented with certain algebraic constraints, which express
connections $\omega_{\alpha(m),\dot\beta(n)}$
via $\frac{1}{2}|m-n|$ derivatives of the dynamical frame-like HS field.
The multispinor \mbox{zero-forms}
$C^{\alpha(2s)}$ and $\bar C^{\dot\beta(2s)}$, which remain non-zero on-shell,
 are spin-$s$ analogues of the Weyl tensor in gravity.

HS gauge transformation is:
\begin{multline}\label{gt}
\delta\omega^{\alpha(m)}{}_{,}{}^{\dot\beta(n)}=
\tilde D\epsilon^{\alpha(m)}{}_{,}{}^{\dot\beta(n)}
+n(\theta(m-n)
+\lambda^2\theta(n-m-2))
\tilde h_{\gamma,}{}^{\dot\beta}\epsilon^{\gamma\alpha(m)}{}_{,}{}^{\dot\beta(n-1)}+\\
+m(\theta(n-m)+
\lambda^2 \theta(m-n-2))
\tilde h^{\alpha}{}_{,\dot\delta}\epsilon^{\alpha(m-1)}{}_{,}{}^{\dot\beta(n)\dot\delta},
\end{multline}
where a gauge parameter $\epsilon^{\alpha(m)}{}_{,}{}^{\dot\beta(n)}(x)$ is an
arbitrary function of $x$.
Note that the limit $\lambda \to 0$ gives the proper description of HS fields
in $4d$ Minkowski space.

As explained in \cite{Smirnov:2013}, to obtain currents with odd and even spins, the connections $\omega^{i;\alpha(m),\dot\beta(n)}$ and curvatures $R^{i;\alpha(m),\dot\beta(n)}$
 should be endowed with a color index
$i = 1 \ldots N$, which labels independent dynamical fields. To contract color indices,
we introduce the real tensor $c_{ijk}$, which can be either symmetric
or antisymmetric. Color indices are raised and lowered by the Euclidean metric $g_{ij}$.
It is convenient to set $g_{ij} = \delta_{ij}$.

Free fields are described by the quadratic action \cite{Vasiliev:1987fortschr}:
\begin{equation}\label{S0}
S_2 = \frac{1}{2} \ \int \sum_{m, n \geq 0} \frac{1}{m!n!} \
\varepsilon(m-n) \ \lambda^{-|m-n|} \
R_1{}^{i;\alpha(m)}{}_{,}{}^{\dot\beta(n)} R_{1 \ i;\alpha(m),\dot\beta(n)},
\end{equation}
where $\varepsilon(x) = \theta(x) - \theta(-x)$ and $m+n=2(s-1)$, $s$ being a spin of the field.

Following \cite{Fradkin:1986qy, Fradkin:1987ks}, to obtain a cubic deformation of the
quadratic action, the linear curvature
$R_1$ in the action (\ref{S0}) has to be replaced by \mbox{$R= R_1 + R_2$} where:
\begin{multline}\label{R2def}
 R_2{}^{i;\alpha(m)}{}_{,}{}^{\dot\beta(n)}
=\sum_{p,q,k,l,u,v \geq 0} \lambda^{1 + d_0 - d_1 - d_2} \
\frac{m!n!}{p!q!k!l!u!v!} \ c^i{}_{jk} \ \delta_{p+q,m}\delta_{u+v,n}
 \times \\
\times\omega^{j;\alpha(p)}{}_{\gamma(k),\dot\delta(l)}{}^{\dot\beta(u)} \
\omega^{k;\alpha(q)\gamma(k)}{}_{,}{}^{\dot\delta(l)\dot\beta(v)},
\end{multline}
\begin{gather*}
d_0 = \frac{|m-n|}{2}, \quad
d_1 = \frac{|p+k-l-u|}{2}, \quad
d_2 = \frac{|q+k-l-v|}{2}.
\end{gather*}

The nonlinear action is:
\begin{equation}\label{S1}
S = \frac{1}{2} \ \int \sum_{m, n \geq 0} \frac{1}{m!n!} \
\varepsilon(m-n) \ \lambda^{-|m-n|} \
R^{i;\alpha(m)}{}_{,}{}^{\dot\beta(n)} R_{i;\alpha(m),\dot\beta(n)}.
\end{equation}

\section{Problem}
It is convenient to describe currents as Hodge-dual differential
forms. The on-shell closure condition for the latter
is traded for the current conservation condition. In this paper, we
consider spin-$t$ currents in $AdS_4$ built from two connections
of integer spins $s_1, s_2 >0$ such that $t \leq s_1 + s_2 - 1$.
Such currents contain the minimal possible number of derivatives of
the dynamical fields.
The analogous problem in $4d$ Minkowski space
has been solved in \cite{Smirnov:2013} for the case of $s_1=s_2$.
The form of the currents will be derived from the nonlinear action (\ref{S1}).

An arbitrary variation of the action (\ref{S1}) can be represented
in the form:
\begin{equation}
\label{do}
\delta S = \int \sum_{t,s_1,s_2} \sum_{m,n} \delta(m + n - 2(t-1)) J_{t,s_1,s_2}{}^{i;\alpha(m)}{}_{,}{}^{\dot\beta(n)} \
\delta \omega_{i;\alpha(m),\dot\beta(n)}\,.
\end{equation}
The current $J_{t,s_1,s_2}{}^{i;\alpha(m)}{}_{,}{}^{\dot\beta(n)}$
carries the color index $i$. Actually, there are $N$ copies of a current,
one for each value of $i$,
and we can set $i=1$ without loss of generality.
In what follows, this index $i=1$ will be omitted in all current forms.
Furthermore, it is convenient to set $c_{jk} := c_{1jk}$ with
$c_{jk}$ being either symmetric or antisymmetric, {i.e.},
\begin{equation}\label{c}
c_{jk} = \eta c_{kj}\,,\qquad \eta^2=1\,.
\end{equation}

To define a nontrivial HS charge as an integral over a $3d$ space, one should find
such a current three-form $J_{t,s_1,s_2}(x)$
built from dynamical HS fields that
is closed by virtue of
HS field Equations (\ref{EM})--(\ref{EMCdot}),
but not exact.
The closed current three-form is:
\begin{equation}\label{xiJ}
J_{t,s_1,s_2}(x)=\sum_{m,n}\frac{\lambda^{-|m-n|}}{m!n!}\delta(m + n - 2(t-1))\xi_{\alpha(m),\dot\beta(n)}(x)
J_{t,s_1,s_2}{}^{\alpha(m)}{}_{,}{}^{\dot\beta(n)} (x),
\end{equation}
where the factor of $\frac{\lambda^{-|m-n|}}{m!n!}$
is introduced for convenience and
$\xi_{\alpha(m),\dot\beta(n)}$ are global symmetry parameters,
which can be identified with those gauge symmetry parameters that leave
the background gauge fields invariant. In accordance with (\ref{gt}),
these parameters obey:
\begin{multline}\label{glob}
D \xi^{\alpha(m)}{}_{,}{}^{\dot\beta(n)}:=
\tilde D\xi^{\alpha(m)}{}_{,}{}^{\dot\beta(n)}
+n(\theta(m-n)+ \lambda^2\theta(n-m-2))
\tilde h_{\gamma,}{}^{\dot\beta}\xi^{\gamma\alpha(m)}{}_{,}{}^{\dot\beta(n-1)}+\\
+m(\theta(n-m)+\lambda^2 \theta(m-n-2))
\tilde h^{\alpha}{}_{,\dot\delta}\xi^{\alpha(m-1)}{}_{,}{}^{\dot\beta(n)\dot\delta}=0.
\end{multline}

One can see that:
\begin{equation}\label{JDD}
d J_{t,s_1,s_2}=
\sum_{m,n}\frac{\lambda^{-|m-n|}}{m!n!}\Big (D\xi_{\alpha(m),\dot\beta(n)}
J_{t,s_1,s_2}{}^{\alpha(m)}{}_{,}{}^{\dot\beta(n)}
+\xi_{\alpha(m),\dot\beta(n)}D J_{t,s_1,s_2}{}^{\alpha(m)}{}_{,}{}^{\dot\beta(n)}\Big)\,.
\end{equation}
Hence, for parameters obeying (\ref{glob}),
the conservation condition amounts to equations:
\begin{equation}\label{ImprConservCond}
D J_{t,s_1,s_2}{}^{\alpha(m)}{}_{,}{}^{\dot\beta(n)} \simeq 0, \qquad m+n=2(t-1).
\end{equation}
For the currents defined via (\ref{do}), the conservation condition (\ref{ImprConservCond})
holds as a consequence of the gauge invariance of the action proven in
\cite{Fradkin:1986qy}.

Conserved currents generate conserved charges. By the Noether theorem, the
latter are generators of global symmetries. Hence, one
should expect as many conserved charges as global symmetry parameters.
For a spin $t$, there are as many global symmetry parameters
as the gauge parameters $\epsilon_{\alpha(m),\dot\beta(n)}$ with \mbox{$m+n=2(t-1)$}.

In what follows, we will use notations:
\begin{equation}\label{Dt}
D^{top}\omega^{\alpha(m)}{}_{,}{}^{\dot\beta(n)} :=
n\theta(m-n)
\tilde h_{\gamma,}{}^{\dot\beta} \ \omega^{\gamma\alpha(m)}{}_{,}{}^{\dot\beta(n-1)}
+m\theta(n-m)\tilde h^{\alpha}{}_{,\dot\delta} \
\omega^{\alpha(m-1)}{}_{,}{}^{\dot\beta(n)\dot\delta},
\end{equation}
\begin{equation}\label{Ds}
D^{sub}\omega^{\alpha(m)}{}_{,}{}^{\dot\beta(n)} :=
n\theta(n-m-2)
\tilde h_{\gamma,}{}^{\dot\beta} \ \omega^{\gamma\alpha(m)}{}_{,}{}^{\dot\beta(n-1)}
+m \theta(m-n-2)
\tilde h^{\alpha}{}_{,\dot\delta} \ \omega^{\alpha(m-1)}{}_{,}{}^{\dot\beta(n)\dot\delta},
\end{equation}
\begin{equation}
\label{Qcur}
D^{cur}\omega^{\alpha(m)}{}_{,}{}^{\dot\beta(n)} :=
R_1{}^{\alpha(m)}{}_{,}{}^{\dot\beta(n)}.
\end{equation}
As a consequence of (\ref{LinHS}):
\begin{equation}\label{DQQ}
D^{cur} = \tilde D + D^{top} + \lambda^2 D^{sub}.
\end{equation}
Since the $\lambda$-dependent term vanishes in the Minkowski case,
it is convenient to introduce the ``flat'' part of the covariant derivative:
\begin{equation}\label{Qfl}
D^{fl} := \tilde D + D^{top}\,.
\end{equation}
It is also convenient to denote:
\begin{equation}
D^{h} := D^{top} + \lambda^2 D^{sub}\,.
\end{equation}

Free field equations (\ref{EM}) imply that:
\begin{equation}\label{QcurC}
D^{cur}\omega^{\alpha(m)}{}_{,}{}^{\dot\beta(n)} \simeq
\delta_{n,0} C^{\alpha(m)\gamma\delta} \
H_{\gamma\delta}+
\delta_{m,0} \bar C^{\dot\beta(n)\dot\gamma\dot\delta} \
\bar H_{\dot\gamma\dot\delta}\,,
\end{equation}
where $\simeq$ implies on-shell equality.

If the three-form $J_{t,s_1,s_2}$ verifies (\ref{ImprConservCond}) on-shell, the charge:
\begin{equation}\label{charge}
 Q_\xi = \int \limits_{M^3} J_{t,s_1,s_2}
\end{equation}
is conserved by virtue of (\ref{glob}). As a result,
there are as many conserved charges $Q_\xi$ as independent global symmetry
parameters $\xi$.
Nontrivial charges are represented by the current
cohomology,
\ie closed currents $J_{t,s_1,s_2} (x)$ modulo exact ones $J_{t,s_1,s_2} \simeq d \Psi_{t,s_1,s_2}$.
Since the currents should be closed on-shell, \ie
by virtue of the free field Equations (\ref{EM})--(\ref{EMCdot}),
analysis is greatly simplified by the fact that
all linearized HS curvatures $R_1{}^{\alpha(m)}{}_{,}{}^{\dot\beta(n)}$ with ${m>0, \ n>0}$
are zero on-shell.

Conservation of currents does not imply that they are invariant under the
gauge transformations (\ref{gt}). However, as shown below,
the gauge variation of $J_{t,s_1,s_2}$ is exact:
\begin{equation}
\delta J_{t,s_1,s_2} (x) \simeq d H_{t,s_1,s_2} (x)
\end{equation}
so that the charge $Q_\xi$ turns out to be gauge invariant.

Thus, the problem is:
\begin{itemize}[leftmargin=*,labelsep=5.8mm]
\item to find current three-forms (\ref{xiJ}) from the variation of action,
\item to check that these forms obey the conservation condition
(\ref{ImprConservCond}),
\item to check that in the flat limit $\lambda \rightarrow 0$, these forms give currents of \cite{Smirnov:2013},
\item to check that the HS charges are gauge invariant.
\end{itemize}

\section{Variation of the Action}
Variation of the nonlinear curvature $R^{i;\alpha(m)}{}_{,}{}^{\dot\beta(n)}$ is:
\begin{equation}
\delta R^{i;\alpha(m)}{}_{,}{}^{\dot\beta(n)} = \delta R_1{}^{i;\alpha(m)}{}_{,}{}^{\dot\beta(n)}
+ \delta R_2{}^{i;\alpha(m)}{}_{,}{}^{\dot\beta(n)},
\end{equation}
where:
\begin{align}\label{deltaR1}
\delta R_1{}^{i;\alpha(m)}{}_{,}{}^{\dot\beta(n)}&=
\tilde D\delta\omega^{i;\alpha(m)}{}_{,}{}^{\dot\beta(n)}+\nonumber\\
&+n(\theta(m-n)
+\lambda^2\theta(n-m-2)) \
\tilde h_{\gamma,}{}^{\dot\beta} \
\delta\omega^{i;\gamma\alpha(m)}{}_{,}{}^{\dot\beta(n-1)}+\nonumber\\
&+m(\theta(n-m)+
\lambda^2 \theta(m-n-2)) \
\tilde h^{\alpha}{}_{,\dot\delta} \
\delta\omega^{i;\alpha(m-1)}{}_{,}{}^{\dot\beta(n)\dot\delta}
\end{align}
and:
\begin{multline}\label{deltaR2}
\delta R_2{}^{i;\alpha(m)}{}_{,}{}^{\dot\beta(n)}
=\sum_{p,q,k,l,u,v}\lambda^{1 + d_0 - d_1 - d_2} \
\frac{m!n!}{p!q!k!l!u!v!}(1+(-1)^{k+l}\eta) \ c^i{}_{jk} \ \delta_{p+q,m}\delta_{u+v,n}
\times\\
\times\delta\omega^{j;\alpha(p)}{}_{\gamma(k),\dot\delta(l)}{}^{\dot\beta(u)} \
\omega^{k;\alpha(q)\gamma(k)}{}_{,}{}^{\dot\delta(l)\dot\beta(v)}
\end{multline}
with $\eta$ defined in (\ref{c}).

Variation of the action (\ref{S1}) is:
\begin{multline}\label{deltaS1}
\delta S
= \int \sum_{m,n} \varepsilon(m-n) \frac{\lambda^{-|m-n|}}{m!n!}
(
R_1{}^{i;\alpha(m)}{}_{,}{}^{\dot\beta(n)} \delta R_{1 \ i;\alpha(m),\dot\beta(n)} +
R_2{}^{i;\alpha(m)}{}_{,}{}^{\dot\beta(n)} \delta R_{1 \ i;\alpha(m),\dot\beta(n)} + \\
+R_1{}^{i;\alpha(m)}{}_{,}{}^{\dot\beta(n)} \delta R_{2 \ i;\alpha(m),\dot\beta(n)}
+R_2{}^{i;\alpha(m)}{}_{,}{}^{\dot\beta(n)} \delta R_{2 \ i;\alpha(m),\dot\beta(n)}
).
\end{multline}
The first term is the variation of the action $S_2$ (\ref{S0}), which
 vanishes on equations of motion (\ref{EM})--(\ref{EMCdot}).
The last term is cubic in connections $\omega^{i;\alpha(m)}{}_{,}{}^{\dot\beta(n)}$,
hence not contributing to bilinear currents.
The second and third terms give rise to the currents.
Using (\ref{LinHS}), (\ref{EMCundot}), (\ref{EMCdot}), (\ref{DQQ}), (\ref{deltaR1}) and (\ref{deltaR2})
and integrating by parts, we obtain:
\begin{multline}\label{finaldelta}
\delta S \simeq
 \int \sum_{m,n} \ \varepsilon(m-n) \ \frac{\lambda^{-|m-n|}}{m!n!} \
[-\tilde D R_2{}^{i;\alpha(m),\dot\beta(n)} \
\delta \omega_{i;\alpha(m),\dot\beta(n)}+\\
+n(\theta(m-n)+\lambda^2\theta(n-m-2)) \
R_2{}^{i;\alpha(m),\dot\theta\dot\beta(n-1)} \
\tilde h^{\gamma}{}_{,\dot\theta} \
\delta\omega_{i;\gamma\alpha(m),\dot\beta(n-1)} \ +\\
+m(\theta(n-m)+ \lambda^2 \theta(m-n-2)) \
R_2{}^{i;\alpha(m-1)\gamma,\dot\beta(n)} \
\tilde h_{\gamma,}{}^{\dot\delta} \
\delta\omega_{i;\alpha(m-1),\dot\beta(n)\dot\delta}
]+\\
+ \int\sum_{r>0} \frac{\lambda^{-r}}{r!} \
(
C^{i;\alpha(r)\gamma\delta} \ H_{\gamma\delta}
\ \delta R_{2 \ i;\alpha(r)}
-\bar C^{i;\dot\beta(r)\dot\gamma\dot\delta} \ \bar H_{\dot\gamma\dot\delta}
\ \delta R_{2 \ i;\dot\beta(r)}
).
\end{multline}

Omitting the color index $i=1$, this leads to the currents at $t > 1$ via:
\begin{equation}
J_{t,s_1,s_2} = \sum_{m,n} \delta(m+n-2(t-1))\xi_{\alpha(m),\dot\beta(n)} \frac{\delta S}{\delta \omega_{\alpha(m),\dot\beta(n)}}\,.
\end{equation}

\section{Examples}
\subsection{Spin-Two Current}
To illustrate the structure of the current three-form and analyze the flat limit $\lambda \rightarrow 0$, consider a current with $t=2$, $s_1=s_2=s > 1$:
\begin{equation}\label{J2}
J_{2,s} = \frac{\lambda^{-2}}{2}\xi_{\alpha\alpha}J_{2,s}{}^{\alpha\alpha}
+ \xi_{\alpha,\dot\beta} J_{2,s}{}^{\alpha}{}_{,}{}^{\dot\beta}
+ \frac{\lambda^{-2}}{2}\xi_{\dot\beta\dot\beta}J_{2,s}{}^{\dot\beta\dot\beta}\,,
\end{equation}
where $J_{2,s} := J_{2,s,s}$\,.
Using (\ref{R2def}), (\ref{Qcur}), (\ref{QcurC}) and (\ref{deltaR2}),
we obtain:
\begin{multline}\label{Jaa}
J_{2,s}{}^{\alpha\alpha}=
\sum_{m,n} \frac{4\lambda^{2-|m-n|} }{(m-1)!n!} c_{ij}[
n(\theta(m-n)+\lambda^2\theta(n-m-2))
\omega^{i; \alpha\gamma(m-1)\varphi}{}_{,}{}^{\dot\delta(n-1)}
\omega^{j;\alpha}{}_{\gamma(m-1),\dot\delta(n-1)\dot\theta}
\tilde h_{\varphi,}{}^{\dot\theta}+\\
+(m-1)(\theta(n-m)+\lambda^2\theta(m-n-2))
\omega^{i; \alpha\gamma(m-2)}{}_{,}{}^{\dot\delta(n)\dot\theta}
\omega^{j;\alpha}{}_{\varphi\gamma(m-2),\dot\delta(n)}
\tilde h^{\varphi}{}_{,\dot\theta}+\\
+(\theta(n-m)+\lambda^2\theta(m-n-2))
\omega^{i; \gamma(m-1)}{}_{,}{}^{\dot\delta(n)\dot\theta}
\omega^{j;\alpha}{}_{\gamma(m-1),\dot\delta(n)}
\tilde h^{\alpha}{}_{,\dot\theta}]\, ,
\end{multline}
\begin{multline}\label{Jab}
J_{2,s}{}^{\alpha}{}_{,}{}^{\dot\beta}=
\sum_{m,n} 2\lambda^{2-|m-n|} [ \frac{1}{(m-1)!n!}c_{ij}
\omega^{i; \alpha\gamma(m-1)}{}_{,}{}^{\dot\delta(n)}
\omega^{j; \varphi}{}_{\gamma(m-1),\dot\delta(n)}
\tilde h_{\varphi,}{}^{\dot\beta}- \\
-\frac{1}{m!(n-1)!}c_{ij}\omega^{i; \gamma(m)}{}_{,}{}^{\dot\delta(n-1)\dot\theta}
\omega^{j; }{}_{\gamma(m),\dot\delta(n-1)}{}^{\dot\beta}
\tilde h^{\alpha,}{}_{\dot\theta}]+\\
+\frac{2\lambda^{4-2s}}{(2s-3)!}c_{ij} [ C^{i;\alpha\gamma(2s-3)\varphi\rho} \
H_{\varphi\rho} \ \omega^{j;}{}_{\gamma(2s-3),}{}^{\dot\beta}
- \bar C^{i;\dot\delta(2s-3)\dot\beta\dot\psi\dot\theta} \
\bar H_{\dot\psi\dot\theta} \ \omega^{j;}{}^{\alpha}{}_{,\dot\delta(2s-3)}]\, ,
\end{multline}
\begin{multline}\label{Jbb}
J_{2,s}{}^{\dot\beta\dot\beta}=
\sum_{m,n} \frac{4\lambda^{2-|m-n|}}{m!(n-1)!} c_{ij}[
m(\theta(n-m)+\lambda^2\theta(m-n-2))
\omega^{i; \gamma(m-1)}{}_{,}{}^{\dot\delta(n-1)\dot\theta\dot\beta}
\omega^{j;}{}_{\varphi\gamma(m-1),\dot\delta(n-1)}{}^{\dot\beta}
\tilde h^{\varphi}{}_{,\dot\theta}+\\
+(n-1)(\theta(m-n)+\lambda^2\theta(n-m-2))
\omega^{i; \varphi\gamma(m)}{}_{,}{}^{\dot\delta(n-2)\dot\beta}
\omega^{j;}{}_{\gamma(m),\dot\delta(n-2)\dot\theta}{}^{\dot\beta}
\tilde h_{\varphi,}{}^{\dot\theta}+\\
+(\theta(m-n)+\lambda^2\theta(n-m-2))
\omega^{i; \varphi\gamma(m)}{}_{,}{}^{\dot\delta(n-1)}
\omega^{j;}{}_{\gamma(m),\dot\delta(n-1)}{}^{\dot\beta}
\tilde h_{\varphi,}{}^{\dot\beta}]\, .
\end{multline}
Recall that $m+n=2(s-1)$, $m,n \geq 0$.

The terms in (\ref{Jaa}), (\ref{Jab}) and (\ref{Jbb}) that contain
inverse powers of $\lambda$ contain higher derivatives.
To obtain a proper $\lambda \to 0$ limit,
such terms should be compensated by an exact form $d\Psi_{2,s}$ with:
\begin{multline}\label{Psi2s}
\Psi_{2,s}
= \sum_{m=0}^{s-3} \frac{2\lambda^{4-2(s-m)}}{m!(2s-3-m!)}
[ \xi_{\alpha\alpha}c_{ij}
\omega^{i;\alpha\gamma(m)}{}_{,}{}^{\dot\delta(2s-3-m)}
\omega^{i;\alpha}{}_{\gamma(m),\dot\delta(2s-3-m)}+\\
+ \xi_{\alpha,\dot\beta} (
 c_{ij}\omega^{i;\alpha\gamma(2s-3-m)}{}_{,}{}^{\dot\delta(m)}
\omega^{j;}{}_{\gamma(2s-3-m),\dot\delta(m)}{}^{\dot\beta}
- c_{ij}\omega^{i; \alpha \gamma(m)}{}_{,}{}^{\dot\delta(2s-3-m)}
\omega^{j;}{}_{\gamma(m),\dot\delta(2s-3-m)}{}^{\dot\beta})-\\
-\xi_{\dot\beta\dot\beta}c_{ij}
\omega^{i;\alpha\gamma(2s-3-m)}{}_{,}{}^{\dot\delta(m)\dot\beta}
\omega^{i;}{}_{\gamma(2s-3-m),\dot\delta(m)}{}^{\dot\beta}]\, .
\end{multline}
At $s=2$, it is not necessary to add this exact form since the current
 is regular in the flat limit.

The fact that complete antisymmetrization over any three two-component dotted
or undotted indices gives zero yields the relation:
\begin{multline}
c_{ij}\omega^{i;\alpha\gamma(m-1)}{}_{,}{}^{\dot\delta(n)}
\omega^{j;\varphi}{}_{\gamma(m-1),\dot\delta(n)} \ \tilde h_{\varphi,}{}^{\dot\beta} =
-c_{ij} \omega^{i;\alpha\gamma(m-1)}{}_{,}{}^{\dot\delta(n-1)\dot\beta}
\omega^{j;\varphi}{}_{\gamma(m-1),\dot\delta(n-1)}{}^{\dot\theta} \ \tilde h_{\varphi,\dot\theta}
+\\
+c_{ij} \omega^{i;\alpha\gamma(m-1)}{}_{,}{}^{\dot\delta(n-1)\dot\theta}
\omega^{j;\varphi}{}_{\gamma(m-1),\dot\delta(n-1)}{}^{\dot\beta} \ \tilde h_{\varphi,\dot\theta}
\end{multline}
to be used in the sequel.

Straightforward calculation gives:
\begin{equation}\label{J2s}
\hat J_{2,s} = J_{2,s} + d\Psi_{2,s} =
\frac{\lambda^{-2}}{2}\xi_{\alpha\alpha}\hat J_{2,s}{}^{\alpha\alpha}
+ \xi_{\alpha,\dot\beta}\hat J_{2,s}{}^{\alpha}{}_{,}{}^{\dot\beta}
+ \frac{\lambda^{-2}}{2}\xi_{\dot\beta\dot\beta}\hat J_{2,s}{}^{\dot\beta\dot\beta},
\end{equation}
where
\begin{multline}\label{adsJaa}
\hat J_{2,s}{}^{\alpha\alpha}=2\lambda^2 c_{ij} [
\omega^{i; \alpha\varphi\gamma(s-2)}{}_{,}{}^{\dot\delta(s-2)}
\omega^{j;\alpha}{}_{\gamma(s-2),\dot\delta(s-2)\dot\theta}\tilde h_{\varphi,}{}^{\dot\theta}+ \\
+ \frac{s-2}{s-1}
\omega^{i; \alpha\gamma(s-3)}{}_{,}{}^{\dot\delta(s-1)\dot\theta}
\omega^{j;\alpha}{}_{\varphi\gamma(s-3),\dot\delta(s-1)}
\tilde h^{\varphi}{}_{,\dot\theta}+\\
+\frac{1}{s-1}
\omega^{i; \alpha\gamma(s-2)}{}_{,}{}^{\dot\delta(s-1)}
\omega^{j;}{}_{\gamma(s-2),\dot\delta(s-1)}{}^{\dot\theta}
\tilde h^{\alpha}{}_{,\dot\theta}]\,,
\end{multline}
\begin{multline}\label{adsJab}
\hat J_{2,s}{}^{\alpha}{}_{,}{}^{\dot\beta}=
\frac{1}{s-1}c_{ij}[
\omega^{i; \gamma(s-2)}{}_{,}{}^{\dot\delta(s-1)\dot\theta}
\omega^{j; }{}_{\gamma(s-2),\dot\delta(s-1)}{}^{\dot\beta}
\tilde h^{\alpha}{}_{,\dot\theta}+\\
+(s-2)\omega^{i; \alpha\gamma(s-3)}{}_{,}{}^{\dot\delta(s-1)\dot\theta}
\omega^{j; }{}_{\varphi\gamma(s-3),\dot\delta(s-1)}{}^{\dot\beta}
\tilde h^{\varphi}{}_{,\dot\theta}+
(s-2)\omega^{i; \alpha\varphi\gamma(s-3)}{}_{,}{}^{\dot\delta(s-1)}
\omega^{j; }{}_{\gamma(s-3),\dot\delta(s-1)}{}^{\dot\theta\dot\beta}
\tilde h_{\varphi,\dot\theta}+\\
+(s-2)\omega^{i; \alpha\gamma(s-1)}{}_{,}{}^{\dot\delta(s-3)\dot\theta}
\omega^{j;\varphi }{}_{\gamma(s-1),\dot\delta(s-3)}{}^{\dot\beta}
\tilde h_{\varphi,\dot\theta}
+(s-2)\omega^{i; \alpha\varphi\gamma(s-1)}{}_{,}{}^{\dot\delta(s-3)}
\omega^{j;}{}_{\gamma(s-1),\dot\delta(s-3)\dot\theta}{}^{\dot\beta}
\tilde h_{\varphi,}{}^{\dot\theta}+\\
+\omega^{i; \alpha\gamma(s-1)}{}_{,}{}^{\dot\delta(s-2)}
\omega^{j;\varphi }{}_{\gamma(s-1),\dot\delta(s-2)}
\tilde h_{\varphi,}{}^{\dot\beta}]\,,
\end{multline}
\begin{multline}\label{adsJbb}
\hat J_{2,s}{}^{\dot\beta\dot\beta} = 2\lambda^2 c_{ij}[
\omega^{i; \varphi\gamma(s-2)}{}_{,}{}^{\dot\delta(s-2)\dot\beta}
\omega^{j;}{}_{\gamma(s-2),\dot\delta(s-2)\dot\theta}{}^{\dot\beta}\tilde h_{\varphi,}{}^{\dot\theta}+\\
+\frac{s-2}{s-1}
\omega^{i;\varphi\gamma(s-1)}{}_{,}{}^{\dot\delta(s-3)\dot\beta}
\omega^{j;}{}_{\gamma(s-1),\dot\delta(s-3)\dot\theta}{}^{\dot\beta}
\tilde h_{\varphi}{}^{\dot\theta}+\\
+\frac{1}{s-1}\omega^{i; \varphi\gamma(s-1)}{}_{,}{}^{\dot\delta(s-2)}
\omega^{j;}{}_{\gamma(s-1),\dot\delta(s-2)}{}^{\dot\beta}
\tilde h_{\varphi,}{}^{\dot\beta}]\,.
\end{multline}
Note that $\hat J_{2,s}$ does not contain $\lambda$ explicitly.
One can check, that
(\ref{adsJaa}), (\ref{adsJab}) and (\ref{adsJbb}) obey
(\ref{ImprConservCond}).
As a result, the form $\hat J_{2,s}$ (\ref{J2s}) is closed by virtue of (\ref{glob}).

Since the
$AdS_4$ current $\hat J_{2,s}$ (\ref{J2s}) does not depend explicitly on $\lambda$
 it preserves its form in the flat limit $\lambda \rightarrow 0$.
From (\ref{adsJaa})--(\ref{adsJbb}), one can see that:
\begin{equation}
\hat J_{2,s}=J_{2,s}^M + D^{fl}\chi_{2,s} ,
\end{equation}
where $J_{2,s}^M$ at $\lambda=0$ reproduces the spin-two current in Minkowski space and:
\begin{multline*}
\chi_{2,s} = \frac{ c_{ij}}{s-1}
\Big ( \xi_{\alpha\dot\beta}\big (\omega^{i; \alpha\gamma(s-2)}{}_{,}{}^{\dot\delta(s-1)}
\omega^{j;}{}_{\gamma(s-2),\dot\delta(s-1)}{}^{\dot\beta}
-\omega^{i; \alpha\gamma(s-1)}{}_{,}{}^{\dot\delta(s-2)}
\omega^{j;}{}_{\gamma(s-1),\dot\delta(s-2)}{}^{\dot\beta}\big )\\
+\lambda^2 \big (
\xi_{\alpha\alpha}
\omega^{i; \alpha\gamma(s-2)}{}_{,}{}^{\dot\delta(s-1)}
\omega^{j;\alpha}{}_{\gamma(s-2),\dot\delta(s-1)}+\xi_{\dot \beta\dot\beta}\omega^{i;\gamma(s-1)}{}_{,}{}^{\dot\delta(s-2)\dot\beta}
\omega^{j;}{}_{\gamma(s-1),\dot\delta(s-2)}{}^{\dot\beta}\big)\Big )
\,.
\end{multline*}
This proves that
the flat limit of the current (\ref{J2s}) reproduces the results of \cite{Smirnov:2013}.

We observe that the current is Hermitian. It is nonzero if $c_{ij}$ is symmetric.

\subsection{Spin-One Current}
Since the action (\ref{S1}) does not contain a kinetic term for spin-one
field $\omega^i$ carrying no spinor indices, following \cite{Fradkin:1986qy},
it should be added separately in a standard way:
\begin{equation}
S_{EM} = \int R_{i}{}^{*} R^i \, ,
\end{equation}
where $^*$ is the Hodge star operator
and, in agreement with (\ref{LinHS}) and (\ref{R2def}),
\begin{equation}
R^{i} = d \omega^i +
\sum_{k,l \geq 0} \
\frac{\lambda^{1 - |m-n|}}{k!l!} \ c^i{}_{jk} \
\omega^{j;}{}_{\gamma(k),\dot\delta(l)} \
\omega^{k;\gamma(k)}{}_{,}{}^{\dot\delta(l)}.
\end{equation}
The full action is $S_{full} = S + S_{EM}$
with $S$ (\ref{S1}).
The spin-one part of the variation
$\delta S_{t=1}$ of this action is:
\begin{equation}\label{deltaSfull}
\delta S_{t=1}
=
\int\sum_{r>0} \frac{\lambda^{-r}}{r!} \
(
C^{i;\alpha(r)\gamma\delta} \ H_{\gamma\delta}
\ \delta R_{2 \ i;\alpha(r)}
-\bar C^{i;\dot\beta(r)\dot\gamma\dot\delta} \ \bar H_{\dot\gamma\dot\delta}
\ \delta R_{2 \ i;\dot\beta(r)}
) + \delta S_{EM},
\end{equation}
where:
\begin{equation}
\delta S_{EM} = \int[R_{1 i}{}^* \delta R_2{}^{i}
+ R_{2i}{}^* \delta R_1{}^{i}].
\end{equation}

In the spin-one case equations,
(\ref{EM})--(\ref{EMCdot}) amount to:
\begin{equation}
R_1{}^{i} = C^{i;\gamma\delta}H_{\gamma\delta}
+\bar C^{i;\dot\gamma\dot\delta}\bar H_{\dot\gamma\dot\delta}\,,
\end{equation}
where $C^{i;\gamma\delta}$ and $\bar C^{i;\dot\gamma\dot\delta}$ parametrize self-dual and anti-self-dual
components of the spin-one field tensor.
Using properties of the Pauli matrices, one can also see that:
\begin{equation}
R_1{}^{i*} = \mathrm{i}(C^{i;\gamma\delta}H_{\gamma\delta}
-\bar C^{i;\dot\gamma\dot\delta}\bar H_{\dot\gamma\dot\delta}).
\end{equation}

The sum of spin-one ($t=1$) currents of fields of
arbitrary spins $s_1=s_2 \geq 1$ can be expressed as
(the color index $i$ is omitted):
\begin{equation}
\xi \frac{\delta S_{t=1}}{\delta\omega} = J_{1,1} + \sum_{s>1} J_{1,s},
\end{equation}
where $\xi$ is a global symmetry parameter zero-form (\ref{glob}) with no spinor indices
and:
\begin{equation}
J_{1,1} =2 \xi\lambda \mathrm{i} c_{ij}
(C^{i;\gamma\delta}H_{\gamma\delta}-\bar C^{i;\dot\gamma\dot\delta}\bar H_{\dot\gamma\dot\delta})\omega^j
- d(\xi R_2{}^{i*}),
\end{equation}
\begin{equation}\label{J1s}
J_{1,s} =
2\lambda^{3-2s}\xi c_{ij}
(
C^{i;\alpha(2s-2)\varphi\rho} \ H_{\varphi\rho} \
\omega^{j;}{}_{\alpha(2s-2)}
-\bar C^{i;\dot\beta(2s-2)\dot\gamma\dot\delta} \ \bar H_{\dot\gamma\dot\delta} \
\omega^{j;}{}_{\dot\beta(2s-2)}\
).
\end{equation}

In the case of $t=1$, $s_1 = s_2 = s=1$,
one can transform $J_{1,1}$ into:
\begin{equation}\label{J11}
\hat J_{1,1} = \frac{1}{\lambda}(J_{1,1} + d(\xi R_2{}^{i*}))
= 2\xi \mathrm{i} c_{ij}
(C^{i;\gamma\delta}H_{\gamma\delta}-
\bar C^{i;\dot\gamma\dot\delta}\bar H_{\dot\gamma\dot\delta})\omega^j.
\end{equation}
It is not hard to see that the $C$-dependent terms are not exact
provided that $c_{ij}$ is antisymmetric.
The current (\ref{J11}) coincides with Minkowski current $J^M_{1,1}$ from \cite{Smirnov:2013}
modulo an overall factor of two.

In the case of $t=1$, $s_1 = s_2 = s > 1$,
the current results from the $C$-dependent terms of (\ref{finaldelta})
by virtue of (\ref{deltaR2}):
\begin{equation}\label{J1s}
J_{1,s} =
2\lambda^{3-2s}\xi c_{ij}
(
C^{i;\alpha(2s-2)\varphi\rho} \ H_{\varphi\rho} \
\omega^{j;}{}_{\alpha(2s-2)}
-\bar C^{i;\dot\beta(2s-2)\dot\gamma\dot\delta} \ \bar H_{\dot\gamma\dot\delta} \
\omega^{j;}{}_{\dot\beta(2s-2)}\
).
\end{equation}
Note, that there are no currents with $t=1$, $s_1 \neq s_2$.
Furthermore, note that Equation (\ref{J11}) is a particular case of (\ref{J1s}) at $s=1$.
The current three-form $J_{1,s}$ (\ref{J1s}) is nontrivial if $c_{ij}$ is
antisymmetric.

For $s>1$, $J_{1,s}$ (\ref{J1s}) can be rewritten in the bilinear form in connections by adding
an exact form:
\begin{multline}
\hat J_{1,s} = -\frac{1}{\lambda (-2)^{s-1} s (s-1)!} (J_{1,s} + d \Psi_{1,s}) =\\
=\xi c_{ij} [\omega^{i; \varphi\gamma(s-2)}{}_{,}{}^{\dot\delta(s-1)}
\omega^{j;}{}_{\gamma(s-2),\dot\delta(s-1)\dot\theta}
+\omega^{i; \varphi\gamma(s-1)}{}_{,}{}^{\dot\delta(s-2)}
\omega^{j;}{}_{\gamma(s-1),\dot\delta(s-2)\dot\theta}
]\tilde h_{\varphi,}{}^{\dot\theta}\,,
\end{multline}
with:
\begin{multline}
\Psi_{1,s}=2\xi\lambda^{3-2s}
\sum_{m=0}^{s-2}(-1)^{m+1} 2^m \lambda^{2m} \frac{(s-1)!}{(s-m-1)!}c_{ij}
(
\omega^{i;\alpha(2s-2-m)}{}_,{}^{\dot\beta(m)}\omega^{j;}{}_{\alpha(2s-2-m),\dot\beta(m)}-\\
-\omega^{i;\alpha(m)}{}_,{}^{\dot\beta(2s-2-m)}\omega^{j;}{}_{\alpha(m),\dot\beta(2s-2-m)}
)\,.
\end{multline}
This three-form is $\lambda$-independent, on-shell-closed, Hermitian
 and reproduces the result of \cite{Smirnov:2013}. Hence, it is
non-trivial.

\section{General Spins}
The $AdS_4$ conserved currents $J_{t,s_1,s_2}$ with $1 < t \leq s_1 + s_2-1$ (for definiteness,
we set $s_1 \geq s_2$)
result from the variation of action (\ref{finaldelta}):
\begin{multline}\label{Jts1s2}
J_{t,s_1,s_2} =
\sum_{m,n} \ \varepsilon(m-n) \ \frac{\lambda^{-|m-n|}}{m!n!} \
[-\xi_{\alpha(m),\dot\beta(n)} \ D^h \
R_2{}^{\alpha(m),\dot\beta(n)}|_{s_1,s_2} - \\
-n(\theta(m-n)+\lambda^2\theta(n-m-2)) \
\xi_{\alpha(m+1),\dot\beta(n-1)} \ R_2{}^{\alpha(m),\dot\theta\dot\beta(n-1)}|_{s_1,s_2} \
\tilde h^{\alpha}{}_{,\dot\theta} + \\
+m(\theta(n-m)+
\lambda^2 \theta(m-n-2)) \
\xi_{\alpha(m-1),\dot\beta(n+1)} \ R_2{}^{\alpha(m-1)\gamma,\dot\beta(n)}|_{s_1,s_2} \
\tilde h_{\gamma,}{}^{\dot\beta}
]+\\
+\sum_{p,q,k,v} \frac{2\lambda^{1-\frac{1}{2}(p+k + |p+q-v| +|q+k-v|)
 }}{p!q!k!v!} c_{ij}\delta_{p+k,2(s_1-1)}
\delta_{q+k+v,2(s_2-1)} \delta_{p+q+v,2(t-1)} \times \\
\big [
 \xi_{\alpha(p+q),\dot\beta(v)}
C^{i;\alpha(p)\gamma(k)\varphi\rho} \ H_{\varphi\rho} \
\omega^{j;\alpha(q)}{}_{\gamma(k),}{}^{\dot\beta(v)}
-  \xi_{\alpha(v),\dot\beta(q+p)}  \bar C^{i;\dot\beta(p)\dot\delta(k)\dot\varphi\dot\rho} \ \bar H_{\dot\varphi\dot\rho} \
 \omega^{j;\alpha(v)}{}_{,\dot\delta(k)}{}^{\dot\beta(q)}\big ],
\end{multline}
where
$R_2{}^{\alpha(m),\dot\beta(n)}|_{s_1,s_2}$ is the restriction of (\ref{R2def})
to terms containing connections with spins $s_1$ and $s_2$.
These currents contain $s_1+s_2-2$ derivatives of the frame-like fields.

To check the non-exactness of the three-form (\ref{Jts1s2}) we further restrict spins to
$1 < t < s_1 + s_2-1$. In this case it suffices to
 add such exact form
\begin{equation*}
d\Psi_{t,s_1,s_2} = d(\sum_{m,n}\xi_{\alpha(m),\dot\beta(n)}
\Psi_{t,s_1,s_2}{}^{\alpha(m),\dot\beta(n)}) \,, \quad n+m=2(t-1)\,,
\end{equation*}
that the resulting current three-form
\begin{equation}
\label{Jhat}
\hat J_{t,s_1,s_2} :=J_{t,s_1,s_2} + d\Psi_{t,s_1,s_2}
\end{equation}
be free of the dependence on the zero-forms $C$.
At $s_1\geq s_2$, the appropriate choice for $\Psi$ in (\ref{Jhat}) is
\begin{multline}
\Psi_{t,s_1,s_2}{}^{\alpha(m),\dot\beta(n)}
=\\
=\sum_{p,q,k,l,u,v} \frac{2\lambda^{1-\frac{|n-m|+|p+k-l-u|+|q+k-l-v|}{2}}}{p!q!k!l!u!v!}
\delta_{p+k+l+u,2(s_1-1)}\delta_{q+k+l+v,2(s_2-1)}
\times \\
\times \theta(l+u-p-k-1)c_{ij}
 [\theta(m-n)\delta_{p+q,m}\delta_{u+v,n}\omega^{i;\alpha(p)\gamma(k),\dot\delta(l)\dot\beta(u)} \
\omega^{j;\alpha(q)}{}_{\gamma(k),\dot\delta(l)}{}^{\dot\beta(v)} - \\
- \theta(n-m) \delta_{p+q,n}\delta_{u+v,m}\omega^{i;\alpha(u)\gamma(l),\dot\delta(k)\dot\beta(p)} \
 \omega^{j;\alpha(v)}{}_{\gamma(l),\dot\delta(k)}{}^{\dot\beta(q)}]\,.
\end{multline}
One can see that $\Psi_{t,s_1,s_2}{}^{\alpha(m),\dot\beta(n)}$ is
adjusted to cancel the $C$-dependent terms.

The resulting current three-form is
\begin{multline}
\hat J_{t,s_1,s_2}{}^{\alpha(m),\dot\beta(n)}
=\\
=\sum_{p,q,k,l,u,v} \frac{2\lambda^{1-\frac{|m-n|+|p+k-l-u|+|q+k-l-v|}{2}}}{p!q!k!l!u!v!}
\delta_{p+k+l+u,2(s_1-1)}\delta_{q+k+l+v,2(s_2-1)} c_{ij}
\times \\
\times [\theta(p+k-l-u-1) D^h
 (\theta(m-n) \delta_{p+q,m}\delta_{u+v,n}
 \omega^{i;\alpha(p)\gamma(k),\dot\delta(l)\dot\beta(u)} \
\omega^{j;\alpha(q)}{}_{\gamma(k),\dot\delta(l)}{}^{\dot\beta(v)} - \\
- \theta(n-m)\delta_{p+q,n}\delta_{u+v,m}\omega^{i;\alpha(u)\gamma(l),\dot\delta(k)\dot\beta(p)} \
 \omega^{j;\alpha(v)}{}_{\gamma(l),\dot\delta(k)}{}^{\dot\beta(q)}) -\\
-\delta_{p+q,m}\delta_{u+v,n}n(\theta(m-n)+\lambda^2\theta(n-m-2))
((u+1) \omega^{i;\alpha(p-1)\gamma(k),\dot\delta(l)\dot\theta\dot\beta(u)} \
 \omega^{j;\alpha(q)}{}_{\gamma(k),\dot\delta(l)}{}^{\dot\beta(v)}
 \tilde h^{\alpha}{}_{,\dot\theta} +\\
+(v+1) \omega^{i;\alpha(p)\gamma(k),\dot\delta(l)\dot\beta(u)} \
 \omega^{j;\alpha(q-1)}{}_{\gamma(k),\dot\delta(l)}{}^{\dot\theta\dot\beta(v)}
 \tilde h^{\alpha}{}_{,\dot\theta} )
+ \\
+\delta_{p+q,m}\delta_{u+v,n}m(\theta(n-m)+
\lambda^2 \theta(m-n-2)) \
((p+1)\omega^{i;\alpha(p)\varphi\gamma(k),\dot\delta(l)\dot\beta(u-1)} \
 \omega^{j;\alpha(q)}{}_{\gamma(k),\dot\delta(l)}{}^{\dot\beta(v)}
\tilde h_{\varphi,}{}^{\dot\beta} + \\
+(q+1)\omega^{i;\alpha(p)\gamma(l),\dot\delta(k)\dot\beta(u)} \
 \omega^{j;\alpha(q)\varphi}{}_{\gamma(l),\dot\delta(k)}{}^{\dot\beta(v-1)}
\tilde h_{\varphi,}{}^{\dot\beta})]\,.
\end{multline}
This current contains $t-|s_1-s_2|$ derivatives, which is
the minimal possible number.
The non-exactness of the current three-form $\hat J_{t,s_1,s_2}$
can be checked in the flat limit $\lambda \to 0$ just as in \cite{Smirnov:2013}.

In the case of $s_1=s_2=s$:
\begin{equation*}
\hat J_{t,s}=\sum_{n,m} \frac{\lambda^{-|m-n|}}{m!n!}\xi_{\alpha(m),\dot\beta(n)}
\hat J_{t,s}{}^{\alpha(m)}{}_{,}{}^{\dot\beta(n)}\,, \quad n+m=2(t-1),
\end{equation*}
where:
\begin{multline}\label{Jts}
\hat J_{t,s}{}^{\alpha(m)}{}_{,}{}^{\dot\beta(n)}=\\
=\lambda^{|m-n|}m!n!(
\theta(n-m-4) \ \hat g(n) \ c_{ij}\omega^{i;\alpha(m)\varphi\gamma(s-2)}{}_{,}{}^{\dot\delta(s-t)\dot\beta(n-t+1)}
\omega^{j;}{}_{\gamma(s-2),\dot\delta(s-t)\dot\theta}{}^{\dot\beta(t-1)}
\tilde h_{\varphi,}{}^{\dot\theta}+\\
+\delta_{n,t} \ c_{ij}[2(t-1)\omega^{i;\alpha(m)\varphi\gamma(s-2)}{}_{,}{}^{\dot\delta(s-t)\dot\beta}
\omega^{j;}{}_{\gamma(s-2),\dot\delta(s-t)\dot\theta}{}^{\dot\beta(n-1)} \tilde h_{\varphi,}{}^{\dot\theta}+\\
+\sum \limits_{p=1}^{t-2} \hat f(p)\omega^{i;\alpha(m)\varphi\gamma(s-p-1)}{}_{,}{}^{\dot\delta(s-t+p)}
\omega^{j;}{}_{\gamma(s-p-1),\dot\delta(s-t+p)}{}^{\dot\beta(n-1)} \tilde h_{\varphi,}{}^{\dot\beta}]+\\
+\delta_{m,t-1}\delta_{n,t-1}c_{ij}
 [\omega^{i; \alpha(t-1)\varphi\gamma(s-2)}{}_{,}{}^{\dot\delta(s-t)}
\omega^{j;}{}_{\gamma(s-2),\dot\delta(s-t)\dot\theta}{}^{\dot\beta(t-1)}
\tilde h_{\varphi,}{}^{\dot\theta}+ \\
+ \omega^{i; \alpha(t-1)\varphi\gamma(s-t)}{}_{,}{}^{\dot\delta(s-2)}
\omega^{j;}{}_{\gamma(s-t),\dot\delta(s-2)\dot\theta}{}^{\dot\beta(t-1)}
\tilde h_{\varphi,}{}^{\dot\theta}] + \\
+
\theta(m-n-4) \ \hat g(m) \ c_{ij}\omega^{i;\alpha(t-1)\varphi\gamma(s-t)}{}_{,}{}^{\dot\delta(s-2)}
\omega^{j;\alpha(m-t+1)}{}_{\gamma(s-t),\dot\delta(s-2)\dot\theta}{}^{\dot\beta(n)}
 \tilde h_{\varphi,}{}^{\dot\theta}+\\
+\delta_{m,t} \ c_{ij}[2(t-1)\omega^{i;\alpha(m-1)\varphi\gamma(s-t)}{}_{,}{}^{\dot\delta(s-2)}
\omega^{j;\alpha}{}_{\gamma(s-t),\dot\delta(s-2)\dot\theta}{}^{\dot\beta(n)} \tilde h_{\varphi,}{}^{\dot\theta}+\\
+\sum \limits_{p=1}^{t-2} \hat f(p)\omega^{i;\alpha(m-1)\gamma(s-t+p)}{}_{,}{}^{\dot\delta(s-p-1)}
\omega^{j;}{}_{\gamma(s-t+p),\dot\delta(s-p-1)}{}^{\dot\theta\dot\beta(n)} \tilde h^{\alpha}{}_{,\dot\theta}]),
\end{multline}
and:
\begin{gather}
\hat g(m) =\frac{2(t-1)!}{(2t-m-2)!(m-t+1)!}\,, \quad m \geq t+1,\\
\hat f(1) = \frac{t - 1}{s-t+1}\,, \quad \hat f(p)=(t-1)\frac{(s-t)!(s-p)!}{(s-3)!(s-t+p)!}\,, \quad p>1.
\end{gather}
The second and last terms in (\ref{Jts})
contribute to the special cases of
$n = t$ and $m = t$, respectively.

One can check that the current (\ref{Jts1s2}) at $s_1=s_2$ reproduces that of
\cite{Smirnov:2013} up to a $D^{fl}$-exact form:
\begin{equation*}
\chi_{t,s} = D^{fl}(\sum_{m,n}\xi_{\alpha(m)\dot\beta(n)}
\chi_{t,s}{}^{\alpha(m),\dot\beta(n)}) \,, \quad n+m=2(t-1),
\end{equation*}
where:
\begin{multline}
\chi_{t,s}{}^{\alpha(m)}{}_{,}{}^{\dot\beta(n)} = \\
=\theta(n-m-2)g(n)c_{ij}\sum \limits_{p=1}^{m+1}
\omega^{i;\alpha(m)\gamma(s-p),\dot\delta(s-t+p-1)\dot\beta(n-t+1)}
\omega^{j;}{}_{\gamma(s-p),\dot\delta(s-t+p-1)}{}^{\dot\beta(t-1)}+\\
+\theta(m-n-2)g(m)c_{ij}\sum \limits_{p=1}^{n+1}
\omega^{i;\alpha(t-1)\gamma(s-t+p-1),\dot\delta(s-p)}
\omega^{j;\alpha(m-t+1)}{}_{\gamma(s-t+p-1),\dot\delta(s-p)}{}^{\dot\beta(n)} + \\
+\delta_{m,t-1}\delta_{n,t-1}fc_{ij}\sum \limits_{p=0}^{[\frac{t}{2}]}
[\omega^{i;\alpha(t-p-1)\gamma(s-2),\dot\delta(s-t+1)\dot\beta(p)}
\omega^{j;\alpha(p)}{}_{\gamma(s-2),\dot\delta(s-t+1)}{}^{\dot\beta(s-t-p)}+\\
+\omega^{i;\alpha(t-p-1)\gamma(s-1),\dot\delta(s-t)\dot\beta(p)}
\omega^{j;\alpha(p)}{}_{\gamma(s-1),\dot\delta(s-t)}{}^{\dot\beta(s-t-p)}],
\end{multline}
where:
\begin{gather}
f=\frac{1}{s-t+1}\,, \qquad g(m) =\frac{t-m}{s-p}\,.
\end{gather}

The conserved currents are nontrivial
 if $c_{ij}$ is antisymmetric for odd $t+s_1+s_2$ and symmetric for even.

Thus, the Hermitian current three-form $\hat J_{t,s_1,s_2}$
is on-shell closed, but not exact. It generates the corresponding real conserved charge
$Q=\int \hat J_{t,s_1,s_2}$ that contains as many symmetry parameters as
local HS gauge symmetries.

\section{Gauge Transformations}
Although the current three-form (\ref{Jts1s2}) is not invariant under the gauge transformations (\ref{gt}),
its gauge variation is exact. Schematically, the proof consists of the following steps.
By virtue of (\ref{DQQ}), the gauge variation of any term
from (\ref{Jts1s2}) $\delta(\xi \omega_1 \omega_2 h)$ can be written as:
\begin{equation*}
\delta(\xi \omega_1 \omega_2 h) = \xi \omega_1 (\tilde D\varepsilon_2
+ D^h\varepsilon_2 )h +
\xi (\tilde D\varepsilon_1 + D^h \varepsilon_1)\omega_2 h.
\end{equation*}
This gives:
\begin{multline*}
\delta(\xi \omega_1 \omega_2 h) =-\tilde D (\xi \omega_1 \varepsilon_2 h - \xi \varepsilon_1 \omega_2 h)
+ \xi (D^{cur} \omega_1) \varepsilon_2 h +
\xi \varepsilon_1 (D^{cur} \omega_2) h + \\
+ (D^h\xi) \omega_1 \varepsilon_2 h
+ \xi (D^h \omega_1) \varepsilon_2 h
-\xi\omega_1 (D^h\varepsilon_2) h
+ (D^h \xi) \varepsilon_1 \omega_2 h +
 \xi \varepsilon_1 (D^h \omega_2) h
- \xi(D^h \varepsilon_1) \omega_2 h.
\end{multline*}
All terms containing $D^{cur}$ are canceled by $\delta (\xi CH \omega h)$.
All terms with $D^h$ cancel each other.

This gives:
\begin{gather*}
\delta J_{t,s_1,s_2} \simeq d H_{t,s_1,s_2},
\end{gather*}
where:
\begin{multline}
H_{t,s_1,s_2} = \\
-\sum_{m,n} \ \varepsilon(m-n)
\sum_{p,q,k,l,u,v}
\delta_{p+q,m}\delta_{u+v,n}\delta_{p+k+l+u,2(s_1-1)}\delta_{q+k+l+v,2(s_2-1)}
c_{ij}\xi_{\alpha(p+q),\dot\beta(u+v)}\times \\
\times
\frac{\lambda^{1-\frac{|m-n|}{2}-\frac{|p+k-l-u|}{2}-\frac{|q+k-l-v|}{2}}}{p!q!k!l!u!v!}
[D^h
\varepsilon^{i;\alpha(p)\gamma(k),\dot\delta(l)\dot\beta(u)} \
\omega^{j;\alpha(q)}{}_{\gamma(k),\dot\delta(l)}{}^{\dot\beta(v)}-\\
-n(\theta(m-n)+\lambda^2\theta(n-m-2)) \
\xi_{\alpha(p+q+1),\dot\beta(u+v-1)}
\times \\
\times(\varepsilon^{i;\alpha(p)\gamma(k),\dot\delta(l)\dot\theta\dot\beta(u)} \
\omega^{j;\alpha(q)}{}_{\gamma(k),\dot\delta(l)}{}^{\dot\beta(v)} \
\tilde h^{\alpha}{}_{,\dot\theta} +
\varepsilon^{i;\alpha(p)\gamma(k),\dot\delta(l)\dot\beta(u)} \
\omega^{j;\alpha(q)}{}_{\gamma(k),\dot\delta(l)}{}^{\dot\theta\dot\beta(v)} \
\tilde h^{\alpha}{}_{,\dot\theta})\\
-m(\theta(n-m)+
\lambda^2 \theta(m-n-2)) \
\xi_{\alpha(p+q-1),\dot\beta(u+v+1)} \times \\
\times(\varepsilon^{i;\alpha(p)\gamma(k),\dot\delta(l)\dot\beta(u)} \
\omega^{j;\alpha(q)}{}_{\gamma(k),\dot\delta(l)}{}^{\dot\beta(v)} \
\tilde h^{\alpha}{}_{,\dot\theta} +
\varepsilon^{i;\alpha(p)\varphi\gamma(k),\dot\delta(l)\dot\beta(u)} \
\omega^{j;\alpha(q)\varphi}{}_{\gamma(k),\dot\delta(l)}{}^{\dot\beta(v)} \
\tilde h_{\varphi,}{}^{\dot\beta})
]+\\
+2\sum_{p,q,k,v} \frac{\lambda^{2-s_1-\frac{|k-p|}{2}}}{p!q!k!v!}
\delta_{2p+q+v,2(t-1)}\delta_{p+k,2(s_1-1)}\delta_{p+q+k+v,2(s_2-1)}
c_{ij} \xi_{\alpha(p+q),\dot\beta(p+v)} \times \\
\times [C^{i;\alpha(p)\gamma(k)\varphi\rho} \ H_{\varphi\rho} \
\varepsilon^{j;\alpha(q)}{}_{\gamma(k),}{}^{\dot\beta(p+v)}
- \bar C^{i;\dot\beta(p)\dot\delta(k)\dot\varphi\dot\rho} \ \bar H_{\dot\varphi\dot\rho} \
 \varepsilon^{j;\alpha(p+q)}{}_{,\dot\delta(k)}{}^{\dot\beta(v)}].
\end{multline}
Details of the derivation of this formula are given in the Appendix for the case of $t=2$, $s_1=s_2=s>1$.

Thus, $\delta J_{t,s_1,s_2}$ is exact on-shell.
The same is true for the spin-one current (\ref{J1s}).
Hence, though the current $J_t$ is not gauge invariant,
the corresponding charge is:
\begin{equation*}
\delta Q_\xi \simeq \int d H_{t,s_1,s_2} = 0.
\end{equation*}

\section{Conclusion}
In this paper, spin-$t$ HS currents $J_{t,s_1,s_2}$ in $AdS_4$,
built from boson fields of arbitrary spins obeying $t \leq s_1 +s_2-1$
are found from the variation principle.
Being represented as three-forms, $J_{t,s_1,s_2}$
are closed, but not exact, hence leading to nontrivial HS charges.
These charges are gauge invariant because $\delta J_{t,s_1,s_2}$ is shown to be exact.

In the $4d$ Minkowski case, in addition to natural parity-even currents, we found ``mysterious''
parity-odd currents \cite{Smirnov:2013}.
In agreement with the conjecture of \cite{Smirnov:2013},
we were not able to extend parity-odd currents to $AdS_4$.
The $\lambda \rightarrow 0$ limit of $\hat J_{t,s}$ (\ref{Jts})
reproduces the parity-even currents of \cite{Smirnov:2013}.

Currents constructed from fields of half-integer spins
can be found analogously.
How to operate with half-integer fields is shown in \cite{Vasiliev:1987fortschr}.
It is important to mention that
for the currents built from fields of half-integer spin,
the computations are essentially different, because Equation (\ref{LinHS}) for half-integer spins contains $\lambda$ instead of $\lambda^2$
\cite{Vasiliev:1987fortschr}.

Let us stress that the derivation of the currents via the action applied
in this paper leads to currents containing the non-minimal number
of derivatives according to \cite{Metsaev:1995re} with the higher-derivative terms corresponding to
certain improvements. This is however anticipated since consistent
cubic HS interactions are known \cite{Fradkin:1986qy} to contain
higher-derivative terms allowing one to preserve HS gauge symmetries
associated with gauge fields of different spins.

\section*{Acknowledgements}

This research was supported in part by the Russian Foundation for Basic Research Grant No 14-02-01172. MV is grateful to Yuri Tatarenko for drawing my attention to incorrect coefficients, indices and spin restrictions in the beginning of section 6 of the original version of this paper.

\section*{Appendix. Example of the Gauge Variation of $J^{t,s}$}
Consider the case of $t=2$, $s_1=s_2=s>1$ (\ref{J2}):
\begin{equation*}
J_{2,s} = \frac{\lambda^{-2}}{2}\xi_{\alpha\alpha}J_{2,s}{}^{\alpha\alpha}
+ \xi_{\alpha,\dot\beta} J_{2,s}{}^{\alpha}{}_{,}{}^{\dot\beta}
+ \frac{\lambda^{-2}}{2}\xi_{\dot\beta\dot\beta}J_{2,s}{}^{\dot\beta\dot\beta}.
\end{equation*}
The gauge variation of (\ref{J2}) under the gauge transformation (\ref{gt}) is:
\begin{equation}\label{deltaJ2}
\delta J_{2,s} = \frac{\lambda^{-2}}{2}\xi_{\alpha\alpha} \delta J_{2,s}{}^{\alpha\alpha}
+ \xi_{\alpha,\dot\beta} \delta J_{2,s}{}^{\alpha}{}_{,}{}^{\dot\beta}
+ \frac{\lambda^{-2}}{2}\xi_{\dot\beta\dot\beta} \delta J_{2,s}{}^{\dot\beta\dot\beta},
\end{equation}
where:
\begin{multline}\label{deltaJaa}
\delta J_{2,s}{}^{\alpha\alpha}= -D^h \Big(
\sum_{m,n} \frac{4\lambda^{2-|m-n|} }{(m-1)!n!} c_{ij}[
\tilde D \varepsilon^{i; \alpha\gamma(m-1)}{}_{,}{}^{\dot\delta(n)}
\omega^{j;\alpha}{}_{\gamma(m-1),\dot\delta(n)} - \\
-D^{top}\varepsilon^{i; \alpha\gamma(m-1)}{}_{,}{}^{\dot\delta(n)}
\omega^{j;\alpha}{}_{\gamma(m-1),\dot\delta(n)} -
\lambda^2 D^{sub}\varepsilon^{i; \alpha\gamma(m-1)}{}_{,}{}^{\dot\delta(n)}
\omega^{j;\alpha}{}_{\gamma(m-1),\dot\delta(n)}]\Big),
\end{multline}
\begin{multline}\label{deltaJab}
\delta J_{2,s}{}^{\alpha}{}_{,}{}^{\dot\beta}=
\sum_{m,n} 2\lambda^{2-|m-n|} [ \frac{1}{(m-1)!n!}c_{ij}
\tilde D \varepsilon^{i; \alpha\gamma(m-1)}{}_{,}{}^{\dot\delta(n)}
\omega^{j; \varphi}{}_{\gamma(m-1),\dot\delta(n)}
\tilde h_{\varphi,}{}^{\dot\beta}- \\
-\frac{1}{(m-1)!n!}c_{ij}
D^{top}\varepsilon^{i; \alpha\gamma(m-1)}{}_{,}{}^{\dot\delta(n)}
\omega^{j; \varphi}{}_{\gamma(m-1),\dot\delta(n)}
\tilde h_{\varphi,}{}^{\dot\beta}- \\
-\frac{\lambda^2}{(m-1)!n!}c_{ij}
D^{sub}\varepsilon^{i; \alpha\gamma(m-1)}{}_{,}{}^{\dot\delta(n)}
\omega^{j; \varphi}{}_{\gamma(m-1),\dot\delta(n)}
\tilde h_{\varphi,}{}^{\dot\beta}- \\
-\frac{1}{m!(n-1)!}c_{ij}\omega^{i; \gamma(m)}{}_{,}{}^{\dot\delta(n-1)\dot\theta}
\tilde D \varepsilon^{j; }{}_{\gamma(m),\dot\delta(n-1)}{}^{\dot\beta}
\tilde h^{\alpha,}{}_{\dot\theta} + \\
+\frac{1}{m!(n-1)!}c_{ij} \omega^{i; \gamma(m)}{}_{,}{}^{\dot\delta(n-1)\dot\theta}
D^{top}\varepsilon^{j; }{}_{\gamma(m),\dot\delta(n-1)}{}^{\dot\beta}
\tilde h^{\alpha,}{}_{\dot\theta}]+\\
+\frac{\lambda^2}{m!(n-1)!}c_{ij}\omega^{i; \gamma(m)}{}_{,}{}^{\dot\delta(n-1)\dot\theta}
D^{sub} \varepsilon^{j; }{}_{\gamma(m),\dot\delta(n-1)}{}^{\dot\beta}
\tilde h^{\alpha,}{}_{\dot\theta}]+\\
+\frac{2\lambda^{4-2s}}{(2s-3)!}c_{ij} [C^{i;\alpha\gamma(2s-3)\varphi\rho} \
H_{\varphi\rho} \ \tilde D \varepsilon_{\gamma(2s-3)}{}^{\dot\beta}
C^{i;\alpha\gamma(2s-3)\varphi\rho} \
H_{\varphi\rho} \ D^{top} \varepsilon_{\gamma(2s-3)}{}^{\dot\beta} -\\
-\bar C^{i;\dot\delta(2s-3)\dot\beta\dot\psi\dot\theta} \
\bar H_{\dot\psi\dot\theta} \ \tilde D \varepsilon^{\alpha}{}_{\dot\delta(2s-3)}
+\bar C^{i;\dot\delta(2s-3)\dot\beta\dot\psi\dot\theta} \
\bar H_{\dot\psi\dot\theta} \ D^{top} \varepsilon^{\alpha}{}_{\dot\delta(2s-3)}]\, ,
\end{multline}
\begin{multline}\label{deltaJbb}
\delta J_{2,s}{}^{\dot\beta\dot\beta}= -D^h \Big(
\sum_{m,n} \frac{4\lambda^{2-|m-n|}}{m!(n-1)!} c_{ij}[
\tilde D \varepsilon^{i; \gamma(m)}{}_{,}{}^{\dot\delta(n-1)\dot\beta}
\omega^{j;}{}_{\gamma(m),\dot\delta(n-1)}{}^{\dot\beta}-\\
-D^{top}\varepsilon^{i; \gamma(m)}{}_{,}{}^{\dot\delta(n-1)\dot\beta}
\omega^{j;}{}_{\gamma(m),\dot\delta(n-1)}{}^{\dot\beta}
-\lambda^2 D^{sub}\varepsilon^{i; \gamma(m)}{}_{,}{}^{\dot\delta(n-1)\dot\beta}
\omega^{j;}{}_{\gamma(m),\dot\delta(n-1)}{}^{\dot\beta}]\Big).
\end{multline}
Rearranging terms in (\ref{deltaJ2}), one can obtain:
\begin{equation*}
\delta J_{t,s} = dH + \chi,
\end{equation*}
where $\chi \simeq 0$ (vanishes on-shell) and $dH$ is exact with:
\begin{multline}
H = -\xi_{\alpha\alpha}D^h\big(
\sum_{m,n} \frac{4\lambda^{2-|m-n|} }{(m-1)!n!} c_{ij}
\epsilon^{i; \alpha\gamma(m-1)}{}_{,}{}^{\dot\delta(n)}
\omega^{j;\alpha}{}_{\gamma(m-1),\dot\delta(n)}\big)+\\
+\sum_{m,n} 2\xi_{\alpha,\dot\beta}
 \frac{\lambda^{2-|m-n|}}{(m-1)!n!}c_{ij}(
\varepsilon^{i; \alpha\gamma(m-1)}{}_{,}{}^{\dot\delta(n)}
\omega^{j; \varphi}{}_{\gamma(m-1),\dot\delta(n)}
\tilde h_{\varphi,}{}^{\dot\beta}+
\omega^{i; \alpha\gamma(m-1)}{}_{,}{}^{\dot\delta(n)}
\varepsilon^{j; \varphi}{}_{\gamma(m-1),\dot\delta(n)}
\tilde h_{\varphi,}{}^{\dot\beta})-\\
-\sum_{m,n} 2\xi_{\alpha,\dot\beta}
\frac{\lambda^{2-|m-n|} }{m!(n-1)!}c_{ij}(\varepsilon^{i; \gamma(m)}{}_{,}{}^{\dot\delta(n-1)\dot\theta}
\omega^{j; }{}_{\gamma(m),\dot\delta(n-1)}{}^{\dot\beta}
\tilde h^{\alpha,}{}_{\dot\theta}-
\omega^{i; \gamma(m)}{}_{,}{}^{\dot\delta(n-1)\dot\theta}
\varepsilon^{j; }{}_{\gamma(m),\dot\delta(n-1)}{}^{\dot\beta}
\tilde h^{\alpha,}{}_{\dot\theta})-\\
-\xi_{\dot\beta\dot\beta}D^h\big (
\sum_{m,n} \frac{4\lambda^{2-|m-n|}}{m!(n-1)!} c_{ij}
\varepsilon^{i; \gamma(m)}{}_{,}{}^{\dot\delta(n-1)\dot\beta}
\omega^{j;}{}_{\gamma(m),\dot\delta(n-1)}{}^{\dot\beta}\big)\,.
\end{multline}

Thus, the on-shell gauge variation of $J_{2,s}$ is exact.

\end{document}